\documentclass[twocolumn]{revtex4}

\usepackage{xcolor}
\usepackage{graphicx}
\usepackage{amssymb}
\usepackage{cancel}
\usepackage{ulem}
\usepackage{lineno}
\usepackage{amsmath}
\usepackage{appendix}
\usepackage{url}
 \usepackage{enumitem}
 \usepackage{hyperref}

\hypersetup{
	colorlinks=true,
	linkcolor=blue,
	citecolor=blue,
	urlcolor=blue
}

\begin{document}

\title
{ Non-normalizable quasi-equilibrium states under fractional dynamics 
}

\author{Lucianno Defaveri$^{1}$, Maike A. F. dos Santos$^{2}$, David A. Kessler$^{1}$, Eli Barkai$^{3}$, Celia Anteneodo$^{2,4}$}

\address{
$^1$Department of Physics, Bar Ilan University, Ramat-Gan
52900, Israel,
$^2$Department of Physics, PUC-Rio,  
Rua Marquês de São Vicente  225, 22451-900 Gávea, Rio de Janeiro, Brazil, 
$^3$Department of Physics, Institute of Nanotechnology and Advanced Materials, Bar Ilan University, Ramat-Gan
52900, Israel,
$^4$National Institute of Science and technology for Complex Systems, 22290-180, Rio de Janeiro, Brazil
}

\begin{abstract}

We study non-normalizable quasi-equilibrium states (NNQE) arising from anomalous diffusion. Initially,  particles in contact with a thermal bath are released from an asymptotically flat potential well, with dynamics that is described by fractional calculus.  
For temperatures that are sufficiently low compared to the potential depth, the properties of the system remain almost constant in time.
We use the fractional-time Fokker-Planck equation (FTFPE) and continuous-time random walk approaches to calculate the ensemble averages of observables. We obtain analytical estimates of the duration of NNQE, depending on the fractional order, from approximate theoretical solutions of the FTFPE. 
 We study and compare two types of observables, the mean square displacement typically used to characterize diffusion, and the thermodynamic energy. We show that the typical time scales for stagnation depend exponentially on the activation energy in units of temperature multiplied by a function of the fractional exponent.


\end{abstract}
 
\maketitle


\section{Introduction}

When Brownian particles are subjected to a confining one-dimensional potential $V(x)$, the probability density function (PDF) $P(x,t)$ of a single particle  attains a normalizable equilibrium state. 
In fact, the Fokker-Planck equation  (FPE)  in one-dimension~\cite{risken} 
\begin{eqnarray} \label{eq:FPE}
\frac{\partial \ }{\partial t} P(x,t)  & = & \mathcal{K}_1  \frac{\partial}{\partial x}\left\{  - \frac{
F(x)}{k_{B} T} + \frac{\partial \ }{\partial x} \right\} P(x,t) \,,
\end{eqnarray}
where   $\mathcal{K}_1$ is the diffusion coefficient, $T$ the temperature, $k_B$ the Boltzmann constant  and $F(x) = - V'(x)$ is the force,  admits the equilibrium solution 
 $P_{\text{eq.}}(x) =Z^{-1}\exp\left[-V(x)/(k_BT)\right]$,  where   $Z=\int_{-\infty}^{\infty}\exp\left[-V(x)/(k_BT)\right]dx$  is the normalizing partition function.

However, if the potential is asymptotically flat, as in Lennard-Jones, Coulomb  and gravitational fields, $Z$ is divergent and this picture breaks down~\cite{fermi,plastino}. 
 We will consider even potentials (i.e., $V(-x)=V(x)$) that behave asymptotically as $V(x) \propto 1/x^\mu$, with $\mu > 1$, specifically of the form
\begin{equation} \label{eq:vmu0}
V(x)=  -\frac{V_0}{\left(1+(x/x_0)^2\right)^{\mu/2}}\,,
\end{equation}
where the length-scale $x_0$ represents the effective region of the potential well. 
For this kind of potential, which  is locally confining,
it has been shown that a kind of quasi-equilibrium state emerges, where  dynamical and thermodynamical observables remain nearly constant ~\cite{Defaveri2020,Anteneodo2021}, with a lifetime that is controlled by the ratio between the temperature and the depth of the potential well. 
 These quasi-equilibrium states are different from those studied by other authors, produced by traps~\cite{bertin2003} or defects~\cite{sollich2005},  related to aging properties.

Although $Z$ is divergent, the average of observables, in these 
so-called non-normalizable quasi-equilibrium (NNQE) states, can be predicted through an appropriate regularization of the  partition  function $Z$. 
 All this is well studied for the FPE (\ref{eq:FPE})~\cite{Defaveri2020,Anteneodo2021}, which describes normal diffusion and relaxation in a force field.  
However,   anomalous diffusion, where the mean square displacement (MSD) scales nonlinearly with time 
as $\left\langle (x-\langle x \rangle)^2 \right\rangle \sim t^{\alpha}$, with $\alpha \neq 1$,  for the force free case,  is ubiquitous in nature, and as such has been reported in many theoretical and experimental works~\cite{Oliveira2019,Bouchaud1990}. 
The MSD was shown to exhibit stagnation, similarly to NNQE in experiments tracking probe particles in micellar solutions, 
with subdiffusive dynamics ~\cite{bellour2002,castillo2008,jeon2013}, in the cytosol of bacteria, yeast, and human cells \cite{Corci2021, Corci2023}. Another example, which exhibits anomalous diffusion, is the complex motion of excitons in semiconductors \cite{Kurilovich2020} and perovskites \cite{Kurilovich2022}, the latter also exhibiting stagnation in the MSD. 
Then, the question emerges: how does anomalous diffusion  affect  the averages and other features  of NNQE states?

To describe the phenomenology of anomalous diffusion,  
different generalizations of the diffusion equation have been proposed, 
with each generalization connected to different mechanisms~\cite{Oliveira2019,Bouchaud1990}, such as, memory effects, non-locality, and disordered or heterogeneous media.
Several methods have been proposed to unravel these mechanisms behind anomalous diffusion in experimental data sets~\cite{Vilk2022,Munoz2021}.
 Three main phenomenological approaches are  nonlinear diffusion~\cite{Tsallis1996,escape2001,nonlinear-book}, fractional Brownian motion~\cite{MandelbrotvanNess}, and the fractional-time  Fokker-Planck approach~\cite{Metzler1999,Metzler2000,Magdziarz2007,Henry2010,Sibatov2013,Sibatov2020,David2022,Metzler2022,Evangelista-book}.  In this manuscript, we will focus on the latter, 
 based on fractional calculus, and identified  as a useful tool  for investigating anomalous diffusion~\cite{Sokolov2001}.  
 In fractional calculus,  different classes of fractional derivatives (integro-differential  operators)  can be defined, with applications in  several  fields of science and engineering~\cite{Podlubny1999}.  
 We  introduce a fractional dynamics by replacing the integer-order temporal derivative in Eq.~(\ref{eq:FPE}) by the fractional Caputo  derivative 
\begin{eqnarray}
  \frac{\partial \ }{\partial t} P  \to  {}^CD_t^{\alpha} P  \equiv  {\displaystyle \frac{1}{\Gamma(1-\alpha)} \int_0^t \frac{1}{(t-t')^{\alpha}}  \frac{d P }{d t'}  dt'} \,,
 \end{eqnarray}
for $0<\alpha<1$,  where $\Gamma(x)$ is the Gamma function. 
This fractional dynamics is particularly interesting as it can be connected with a continuous time random walk  (CTRW) description. 
In the limit  $\alpha\to 1$, the integer-order partial derivative is recovered. 
Then, the fractional time FPE (FTFPE), which generalizes Eq.~(\ref{eq:FPE}), is 
 \begin{eqnarray}
  {}^{C}D_t^{\alpha} P(x,t) & = & \mathcal{K}_{\alpha} \frac{\partial}{\partial x}\left\{  - \frac{
F(x)}{k_{B} T} + \frac{\partial \ }{\partial x} \right\} P(x,t) , \label{eq:fractional-FP}
 \end{eqnarray}
where $\mathcal{K}_{\alpha}$
 is the generalized diffusion coefficient. 
 Alternatively, it can be rewritten   as~\cite{Barkai2000,Podlubny1999}  
\begin{equation} \label{eq:FTFP}
\frac{\partial \ }{\partial t} P(x,t)  = \mathcal{K}_{\alpha}\;{}^{RL}D_t^{1-\alpha}   \frac{\partial}{\partial x}\left\{  - \frac{
F(x)}{k_{B} T} + \frac{\partial \ }{\partial x} \right\} P(x,t) , 
\end{equation}
where
\begin{equation} \label{eq:fderivative}
 {}^{RL}D_t^{1-\alpha}\psi(t)   \equiv  {\displaystyle \frac{1}{\Gamma(\alpha)} \frac{d \ }{d t}  \int_0^t \frac{\psi(t')}{(t-t')^{1-\alpha}} dt'}  
\end{equation}
is the  fractional derivative of a function $\psi(t)$ in the Riemann-Liouville sense.  
In the free case ($F(x)=0$),  Eq.~(\ref{eq:FTFP})  produces subdiffusion as $\left\langle (x-\langle x \rangle)^2 \right\rangle \sim t^{\alpha}$~\cite{Metzler2000}
with an exponent that coincides with the fractional order $\alpha$. 
For the class of potentials of interest,  which are asymptotically flat,  nearly free subdiffusion  occurs at large distances. 

 We complete the model given by Eq.~(\ref{eq:FTFP})  
using the family of asymptotically flat potentials in Eq.~(\ref{eq:vmu0}).
Re-scaling  
\begin{equation} \label{eq:scaling}
x/x_0 \to x, \;\;\; V(x)/V_0 \to v(x), \;\;\;t/t_0 \to t, 
\end{equation}
with $t_0= (x_0^2/\mathcal{K}_{\alpha})^{1/\alpha}$,
and defining $\xi=k_B T/V_0$,  we finally obtain the dimensionless form of Eq.~(\ref{eq:FTFP}), namely,
\begin{eqnarray} \label{eq:FTFPs}
\frac{\partial \ }{\partial t } P(x,t)  & = &   {}^{RL}D_t^{1-\alpha} \frac{\partial}{\partial x}\left\{  - \frac{
f(x)}{\xi} + \frac{\partial \ }{\partial x} \right\} P(x,t), \;\;\;\;\;
\end{eqnarray}
where $f(x)=-v'$, with
\begin{equation} \label{eq:vmu}
v(x) =  -\frac{1}{\left(1+x^2\right)^{\mu/2}}.
\end{equation}

\begin{figure}[t!]  
\centering
\includegraphics[width=0.5\textwidth]{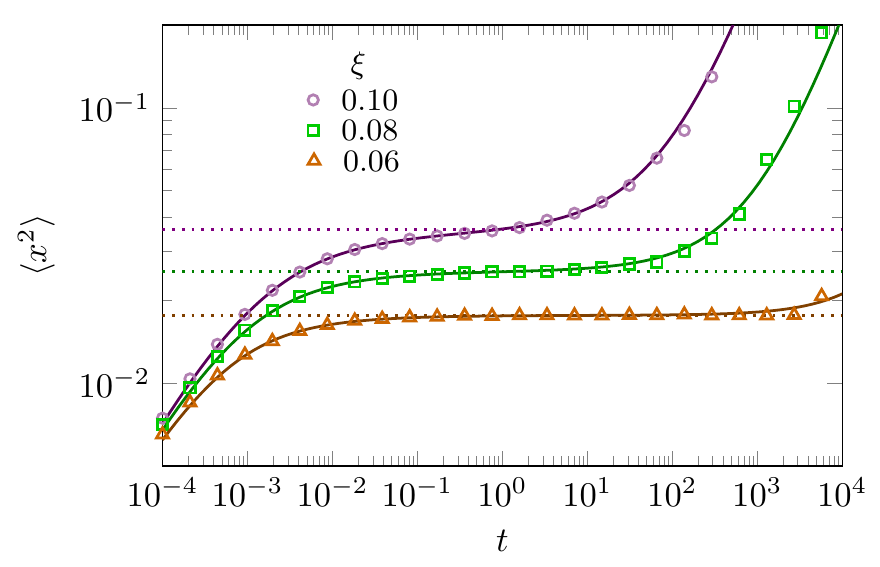}  
\caption{ 
 MSD vs. time, different values of $\xi$ indicated in the legend. 
In all cases, the fractional order is $\alpha=0.6$, and the decay of the potential is ruled by  $\mu=4$. 
Solid lines represent the exact result obtained from the numerical integration of the FTFPE~(\ref{eq:FTFPs}), 
as described in Sec.~\ref{sec:solving}.
Symbols correspond to results obtained from CTRW simulations, 
over $5\times10^5$ trajectories (symbols), 
using $\sigma=10^{-2}$ and $\tau=\{\sigma^2/(2\Gamma[1-\alpha])\}^{1/\alpha}\simeq 1.8\times10^{-8}$ (see Sec.~\ref{sec:CTRW}). 
The NNEQ level, predicted by Eq.~(\ref{eq:xc}),  is plotted by dotted lines. 
}
\label{fig:CTRW}
\end{figure}

Initially the particles are placed at the minimum of the potential, and we expect that for long times they will diffuse. 
 In Fig.~\ref{fig:CTRW} we illustrate the behavior of the MSD as a function of time, for $\alpha=0.6$ and different values of $\xi$, where full lines were obtained by numerical integration of the FTFPE for the initial condition $P(x,0)=\delta(x)$.  
For short times, the packet initially localized at the origin (sub)diffuses almost freely, with the MSD growing as $\langle x^2 \rangle \sim t^\alpha$.
A plateau emerges, for sufficiently low $\xi$, signaling a NNQE where the MSD becomes stagnated for some time. 
The duration of the NNQE increases with $1/\xi$, with its end due to the escape of the particles.
Clearly, beyond the MSD, different observables will have plateaus with different lifespans.

Our purpose is to characterize the NNQE states that emerge at some length and time scales for the class of potentials (\ref{eq:vmu}), 
with a well at the origin and  asymptotically flat, under the chosen fractional dynamics. 
For example, what are the values of the MSD or the energy of the system in stagnation, namely during the NNQE time span?

 Let us mention that escape  problems in the scenario of fractional dynamics have been explored before for different potential and boundary conditions~\cite{MetzlerKlafter2000,Metzler2000,Dybiec2015,Sliusarenko2010}, but not for flat potentials as those here considered.  
 The flatness of the attractive potential is important as it makes it easier for some particles to return to the well, 
 unlike the typical escape problem where the particle must overcome a potential barrier and, once it escapes the well, is repelled away. 
 Therefore, we  deal with a  phenomenon that has not been treated before. 
  
The remaining of the manuscript is organized as follows. 
In Sec.~\ref{sec:solving}, we obtain the solution of the fractional problem, 
focusing on the intermediate long-time regime, corresponding to NNQE states. 
In Sec.~\ref{sec:CTRW}, we present results for the microscopic 
counterpart of the FTFPE  provided by a CTRW, 
in good agreement with the  FTFPE description. 
The impact of the fractional order $\alpha$ on the NQE states is studied in Sec.~\ref{sec:times},
for the mean square displacement (MSD)  and  the energy, as examples of dynamic and  thermodynamic observables, respectively. 
Sec.~\ref{sec:final} contains final remarks.

\section{Solving the fractional-time FPE}
\label{sec:solving}

For the initial condition $P(x,0)=\delta(x)$, the solution of the FTFPE is expected to  present  three main distinct temporal regimes, as we observed in Fig.~\ref{fig:CTRW}.
 First, for very short times, subdiffusion is nearly free at the bottom of the well, and anomalous, with the MSD increasing as $\langle x^2 \rangle \sim t^\alpha$. 
 Second, at intermediate long-time scales, the NNQE state occurs.
  Third, at very long times,  when most of the particles have escaped the well, anomalous diffusion with exponent $\alpha$ should  be predominant again, since 
  $P(x,t)$ will be dominated by the   $x \gg 1$ limit, where  $v(x) \to 0$.

Although it is possible to derive the solutions of the FTFPE rigorously using an eigenfunction expansion, analogously to what was done for $\alpha=1$~\cite{Anteneodo2021}, here we will focus on a much simpler procedure that maps the already known $\alpha=1$ solution onto the fractional problem.

\subsection{Mapping procedure}
\label{sec:mapping}

 The solution of the FTFPE,  $P_\alpha(x,t)$, for any $0<\alpha\le 1$,
 for the initial condition $P_\alpha(x,0)=\delta(x)$,  
 can be expressed, using a subordination technique, as used by several authors in similar contexts~\cite{Gorenflo2007, Weron2010,Ewa2010, Wang2020,Chechkin2021,Wang2022,Zhou2022}.  
 Namely, we use  the following transform of the solution for the non-fractional case ($\alpha=1$)~\cite{Barkai2001}  
\begin{eqnarray}
 P_\alpha(x,t) & = & \int_0^{\infty} n_{\alpha}(q,t)P_1(x,q) dq\,, \label{eq:mapping} \end{eqnarray}
 with
\begin{eqnarray} \label{eq:nalfa}
 n_{\alpha}(q,t) = \frac{t}{\alpha\; q^{1+1/\alpha}}
 L_{\alpha}\left( t/ q^{1/\alpha} \right),
\end{eqnarray}
where $L_\beta(z)$ is the one-sided Lévy PDF, whose Laplace $z\to s$ transform is $\tilde{L}_\beta=\exp(-s^\beta)$~\cite{penson}.

 \begin{figure}[t!]  
\centering
\includegraphics[width=0.45\textwidth]{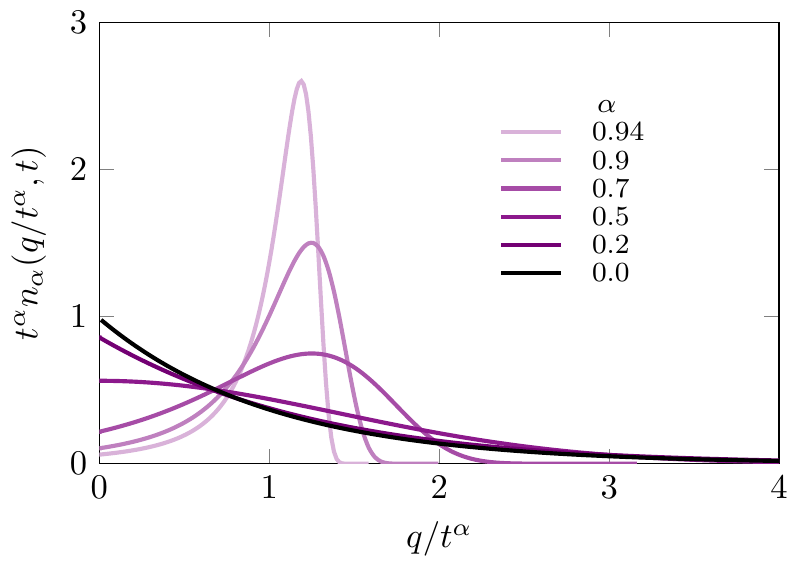}
\caption{
 Transformation function $n_\alpha$ vs. $q/t^\alpha$, scaled such that curves for different times collapse, according to Eq.~(\ref{eq:nalfa}), for different values of $\alpha$ 
(increasing from darker to lighter).   
}
\label{fig:eta}
\end{figure}

Within the framework of  continuous time random walks (CTRW), the function $n_\alpha(q,t)$ can be interpreted as the probability of $q$ steps in the interval $(0,t)$~\cite{Barkai2001}. 
When $\alpha \to 1$, $n_\alpha(q,t) \to \delta(q-t)$. In the opposite case $\alpha\to 0$, we have   $n_\alpha(q,t)\to \exp(-q)$, independently of $t$, implying  
that the initial density does not evolve. 
For intermediate values of $\alpha$, 
the shape of $n_\alpha(q,t)$ is represented in Fig.~\ref{fig:eta}, in terms of the scaled variable $q/t^\alpha$, for which the curves for any $t$ collapse, as it can be straightforwardly derived from Eq.~(\ref{eq:nalfa}). In particular, for $\alpha=1/2$, $n_{1/2}(q,t)$ corresponds to half a Gaussian.

Some succeeding calculations can be simplified by using the Laplace transform of Eq.~(\ref{eq:mapping}), namely,
\begin{eqnarray} \label{eq:mappingLAPLACE}
 \Tilde{P}_\alpha(x,s) & = & \int_0^{\infty} \Tilde{n}_{\alpha}(q,s) P_1(x,q)dq \nonumber \\
 & = & \int_0^{\infty} s^{\alpha-1} e^{-s^{\alpha} q } P_1(x,q) dq \nonumber \\
 & = & s^{\alpha-1}  \Tilde{P}_1(x,s^{\alpha}) \,.
\end{eqnarray}
Then,  the PDF is obtained by inversion of the Laplace transform, namely,  
\begin{equation} \label{eq:Ltrans}
   P_\alpha(x,t)  =  \mathcal{L}^{-1}\{  s^{\alpha-1}  \Tilde{P}_1(x,s^{\alpha})\} \,.
\end{equation} 

Furthermore, the mapping (\ref{eq:mapping})  can be applied  straightforwardly to averaged quantities. Namely, for an observable $O$, we have
\begin{eqnarray} \label{eq:Oalfa}
\langle O \rangle_\alpha & = & \int_0^{\infty} n_{\alpha}(q,t) \langle O \rangle_{1}(q) dq \,,
\end{eqnarray}
where $\langle O \rangle_1 = \int_{-\infty}^{\infty} O P_1(x,t)dx$.

\subsection{Solution for the integer-order FPE ($\alpha=1$)} \label{sec:sol-1}

For the FPE (\ref{eq:FPE}) with integer-order time derivative ($\alpha=1$), an approximate solution was found in previous work~\cite{Anteneodo2021},   based on the eigenfunction expansion of the time-dependent solution of the FPE, with free boundary conditions.
Let us summarize the approximate solution found
for intermediate timescales, where time-independence emerges, and for potentials that decay with exponent $\mu>1$. 

For the central region ({\bf C}), 
 $x \ll t^\frac{1}{2}$, 
\begin{eqnarray}
P^{\text{\textbf{C}}}_1(x,t) 
& \simeq &    \frac{e^{-v(x)/\xi}}{\mathcal{Z}} + {\cal O}(t^{-1/2}), \label{eq:Central1}
\end{eqnarray}
while, for the region  of the tails ({\bf T}),  $x\gg t^\frac{1}{2}$,  
\begin{eqnarray}  
 P^{\text{\textbf{T}}}_1(x,t) & \simeq &  \frac{1}{\mathcal{Z}} \text{erfc}\left( \frac{|x| }{2\sqrt{t}} \right), \label{eq:P3usual}
\end{eqnarray}
 where  
\begin{eqnarray} \label{eq:ell0}
\mathcal{Z} &\equiv&   
2\int_0^\infty (e^{-v(x)/\xi}-1)dx 
 \approx  \sqrt{\frac{2\pi \xi}{\mu}} \;e^{1/\xi}, 
\end{eqnarray}
which plays the role of a regularized BG partition function, 
 does not depend on $\alpha$, and its approximate value was obtained for small $\xi$~\cite{Anteneodo2021}.

 Note that, as expected, the central region is dominated by the Boltzmann factor.  
 However, since it is non-normalizable, Eq.~(\ref{eq:Central1}) is only valid for a
 range of positions inside and near the well (that is, $x \sim O(1)$). 
 Meanwhile,  for large values of $x$, the tails are given by the complementary error function erfc that governs the free diffusion, as the force in that region is negligible, of the very small number of  particles that have managed to escape the deep well into the intermediate regime.

\subsection{Solution for the fractional-time FPE ($0<\alpha<1$)} \label{sec:sol-alfa}

Let us use the method described at the beginning of  subsection~\ref{sec:mapping} to obtain the PDF $ {P}_\alpha(x,t) $ in regions  {\bf C} and {\bf T}. 

For region {\bf C}, $x\ll t^{\alpha/2}$, using Eq.~(\ref{eq:Ltrans}), we have
\begin{eqnarray}
P_\alpha^{\text{\textbf{C}}}(x,t)  & = & \mathcal{L}^{-1} \left\{ s^{\alpha-1} \Tilde{P}^{\text{\textbf{C}}}_1(x,s^{\alpha})    \right\} \nonumber \\
 & \simeq & \mathcal{L}^{-1} \left\{ \frac{e^{- v(x)/\xi}}{\mathcal{Z}} \frac{1}{s} \right\} +{\cal O}(t^{-\alpha/2})\nonumber \\
  & \simeq &   \frac{e^{- v(x)/\xi}}{\mathcal{Z}}  +{\cal O}(t^{-\alpha/2}).  \label{eq:P1alfa}
\end{eqnarray}
Let us highlight that,   for sufficiently long times  (but not so long that the tails of the distribution dominate),  this central region of the PDF remains time-independent, independently of $\alpha$, with the BG shape normalized by the regularized partition function $\mathcal{Z}$.
 
\begin{figure}[t!]    
\includegraphics[width=0.48\textwidth]{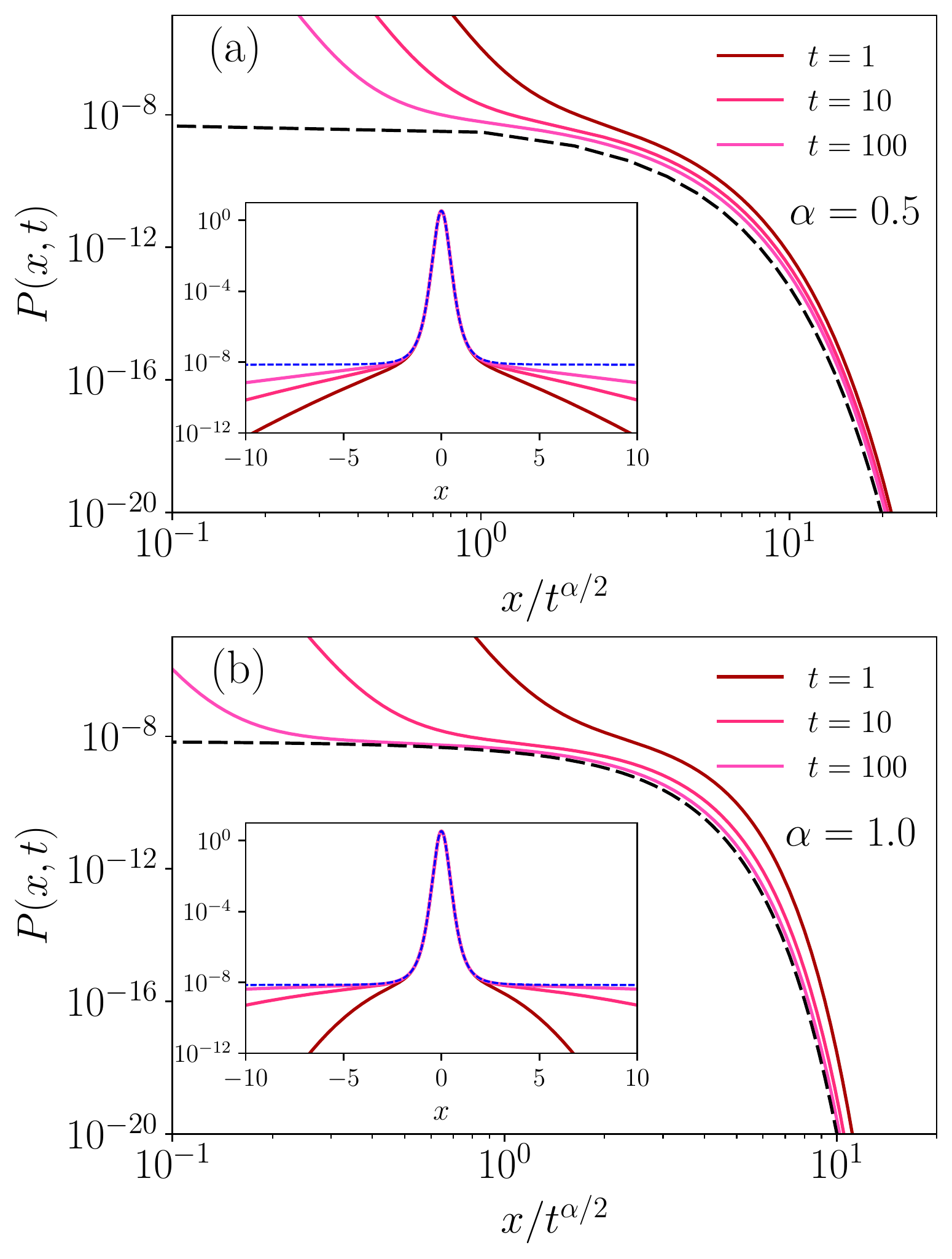} 
\caption{  
Numerical solution of the FTFPE (solid lines) at different times $t$ for $\alpha=$ 0.5 (a) and 1.0 (b). 
The inset shows the exact solutions around the central region (solid lines) at different $t$, together with the  Boltzmann 
factor $e^{-v(x)/\xi}/\mathcal{Z}$ (blue dotted line), which dominates
the approximate the solution for region {\bf C},  given by Eq.~(\ref{eq:P1alfa}). In the main plot, we highlight the tails (only the positive semi-axis), together with the   approximate solution  for region {\bf T}, given by Eq.~(\ref{eq:P3alfa}), in dashed lines. 
The potential has exponent $\mu=4$, and the relative temperature is $\xi=0.05$.
}
\label{fig:Pxt24}
\end{figure}

Analogously, for region {\bf T},  $x\gg t^{\alpha/2}$, 
\begin{eqnarray}
P_\alpha^{\text{\textbf{T}}}(x,t)  & \simeq   & 
\mathcal{L}^{-1}\left\{ s^{\alpha-1} \Tilde{P}^{\text{\textbf{T}}}_1(x,s^{\alpha})  \right\}
\nonumber \\
& = &\mathcal{L}^{-1}\left\{ \frac{1}{\mathcal{Z} s }  e^{-s^{\frac{\alpha}{2}} |x| } \right\}  
\nonumber \\
  & = &    
  \frac{\int_0^t L_\frac{\alpha}{2}
  \left(t'/|x|^{\frac{2}{\alpha}}\right)  dt'}
  {\mathcal{Z}\; |x|^\frac{2}{\alpha}}\,. 
  \label{eq:P3alfa}
\end{eqnarray}
Since $L_\beta$ is the one-sided L\'evy density, the integral is the corresponding cumulative distribution.  
 This expression represents a generalization of the complementary error function, decaying for large $|x|$.
The crossover position $\ell$ between both regions, {\bf C} and {\bf T}, scales as $\ell \sim t^{\alpha/2}$,  due to subdiffusion. 
In Fig.~\ref{fig:Pxt24}(a), we plot the exact numerical solution (solid lines) 
for $\alpha=0.5$ at different times $t$, with the central region matching the Boltzmann factor (inset) and the tails in the main plot approaching the  region  {\bf T} solution for larger $x$ and $t$. For comparison we also show the results for $\alpha=1$ in panel (b). Actually, there is a shift that can be exactly calculated~\cite{Anteneodo2021} producing a perfect match of the tails, but for simplicity we neglected it as it does not affect the physical conclusions 
regarding NNQE values and scaling laws.
Although other procedures exist~\cite{Deng2007,Deng2009},   the numerical solution of the FTFPE was obtained   through the mapping procedure given by Eq.~(\ref{eq:mapping}), using the numerical solutions for the integer order problem ($\alpha = 1$) obtained using a standard Crank-Nicolson integration scheme. 
 
Let us find the condition under which
most of the probability is concentrated in the time-independent region {\bf C}. 
To find out the time at which this assumption fails, we calculate how much probability has flowed from region {\bf C} to region {\bf T} by a given time $t$. By considering a point $x=\ell  $ in the overlap interval, the whole probability in region {\bf T}, from  Eq.~(\ref{eq:P3alfa}), scales as  
\begin{eqnarray}
 \int_{\ell}^{\infty} P_\alpha^{\textbf{T}} (x,t) dx   & =&  \mathcal{L}^{-1}\left\{ \frac{1}{\mathcal{Z} s }\int_{\ell}^{\infty}   e^{-s^{\frac{\alpha}{2}} |x|} dx \right\}
\nonumber \\
 &  =   & \frac{t^{\frac{\alpha}{2}}}{\mathcal{Z} \Gamma(\frac{\alpha}{2} +1)}  
 +{\cal O}(1), \label{eq:tail-correction}
\end{eqnarray}
which recovers the known result when  $\alpha=1$. 
The time dependence is negligible for 
 small values of $\int_{\ell}^{\infty} P_\alpha^{\textbf{T}}(x,t) dx$, which occurs 
 for times such that
\begin{eqnarray} \label{eq:t*}
t  \ll t^*  \equiv
\left\{ \mathcal{Z} \,\Gamma\left(\frac{\alpha}{2} +1 \right) \right\} ^\frac{2}{\alpha}   \sim  
e^\frac{2}{\alpha\xi}\,.
\end{eqnarray}
The scaling of $t^{*}$ quantifies how decreasing $\alpha$ prolongs the departure from the time-independent regime. 
 Note that the scaling exponent is different from that of the escape time  to overcome a potential barrier, with a repelling force driving the particle away from the well at large distances,  namely 
$ t \sim  e^{1/(\alpha \xi)}$~\cite{Metzler2000}.
 The information on the shape of the potential is embodied in $\mathcal{Z}$, given by Eq.~(\ref{eq:ell0}), implying that the more flat the potential (i.e.,  the larger is $\mu$), the window of time-independence shortens,  which is expected as for larger values of $\mu$ the force decays faster and becomes negligible at shorter distances.

\section{CTRW approach}
 \label{sec:CTRW}

A microscopic counterpart of  the FTFPE (\ref{eq:FTFPs}) can be simulated using a continuous time random walk  (CTRW)~\cite{Barkai2000}. 
 In this process, a particle (the walker) moves through a one-dimensional lattice by jumping either left or right with a given probability. 
The waiting times between consecutive jumps are chosen from a probability density distribution given by $\psi(t) = \alpha \tau^{-1} (\tau/t)^{\alpha+1}$, for $t>\tau$, and zero otherwise. 
The available positions for the walker on the lattice are $x = j\sigma$, where $j$ is an integer. 
At time $t=t_n=\sum_{i=1}^{n} t_i$, with $t_i$ drawn from $\psi(t_i)$,  the position of the random walker is given by  $x_{n} = x_{n-1}  \pm \sigma$,  that is, the walker jumps from a site $j$ to site $j\pm 1$, which occurs with probability 
\begin{equation} \label{eq:pj}
p_j = \frac{1}{2}(1 \pm g_j) \,,
\end{equation}
where $g_j$ (such that $|g_j| \le 1$) is given by~\cite{Bel2006}
\begin{eqnarray}
    g_j & = &  \frac{\sigma}{2 \xi} f( j\sigma  ) = 
    - \frac{\mu  }{2 \xi  }\frac{j \sigma^2  }{\left( 1 +  ( j\sigma )^2  \right)^{\frac{\mu}{2}+1 }}.
    \label{eq:gj} 
    \nonumber
\end{eqnarray}
In Eq. (\ref{eq:pj}), we have used thermal detailed balance and hence the jump probabilities depend on temperature as usual. 
Note that small enough $\sigma$, suitable for the continuum FTFPE~\cite{Bel2006},  requires $|g_j| \ll 1$.
The bias $g_j$ renders the jump probability space-dependent, emulating the effect of the potential, and vanishes under  free subdiffusion where 
the walker moves to the left or to the right with equal probability $1/2$. 
Let us note that in the free case, Eq.~(\ref{eq:FTFPs}) yields   $\langle x^2\rangle= (2/\Gamma[\alpha+1]) t^\alpha$,
when  $0< \alpha < 1$.
Figure~\ref{fig:CTRW} displays results obtained from simulations of trajectories  based on the CTRW model for $\alpha=0.6$ (symbols), for a  packet  of particles starting at the origin. 
 These results are in good agreement with the solid lines obtained from the numerical integration of the FTFPE, described in Sec.~\ref{sec:solving}. 

To establish a connection between the microscopic description and  the FTFPE, we must have $\sigma \to 0$ and time $\tau\to 0$  such that $K_\alpha$ remains well defined  according to the rescaling in Eq.~(\ref{eq:scaling}).
This equation is analogous to setting
the generalized diffusion constant $K_\alpha=1$, 
which in turn means that $\sigma$ and $\tau$ are not independent, and we have  
$\sigma^2= 2\Gamma[1-\alpha]\tau^\alpha$~\cite{Barkai2000}.
 
 The effects of the fractional order $\alpha$ on  the NNQE states will be shown in the next sections based on the analytical results, as well as numerical integration of the FTFPE, which are less time-demanding than simulations of numerical trajectories. 
 
\section{NNQE lifespan}
\label{sec:times}

 In this section, we analyze the effect of the fractional order $\alpha$ on the characteristics of quasi-equilibibrium, taking the MSD and the energy of the particle, as representative dynamical and thermodynamical quantities, respectively.

\subsection{Mean-square displacement}
\label{sec:MSD}

\begin{figure}[t!]  
\centering
\includegraphics[width=0.5\textwidth]{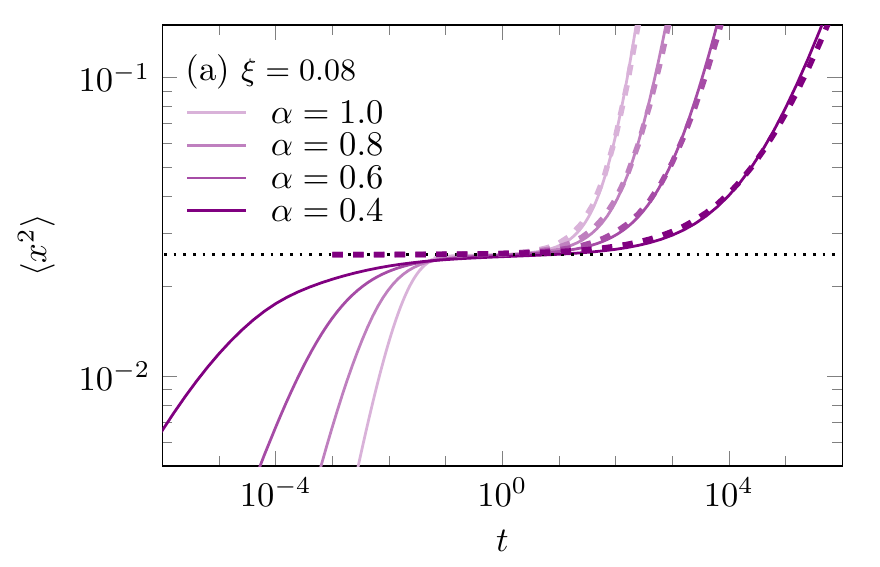}  
\includegraphics[width=0.5\textwidth]{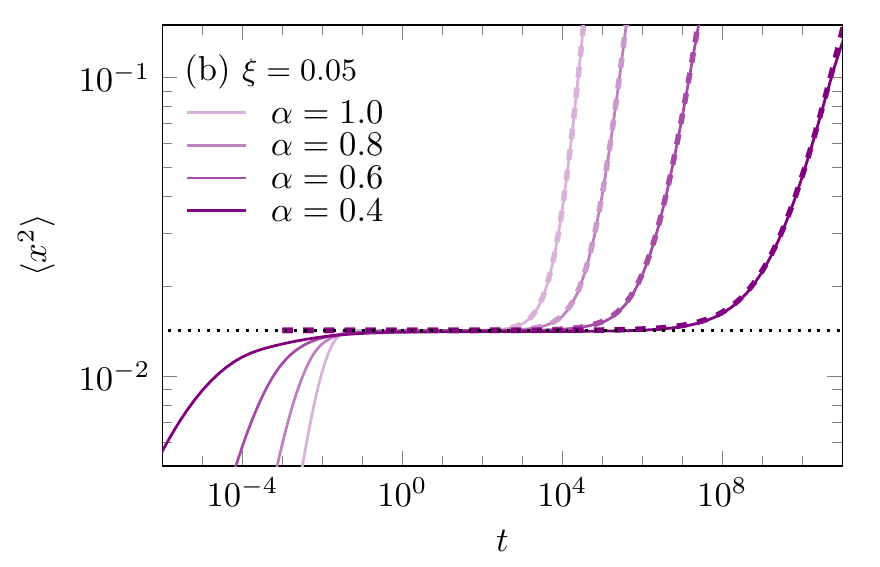}
\caption{MSD vs. time for a potential with $\mu=4$, being
  $\xi=$  0.08 (a) and 0.05 (b).
The  solid lines are the exact numerical solution for the fractional cases.
 Dashed lines represent the averages performed over the approximate PDFs, according to  Eq.~(\ref{eq:MSD}). 
 Notice that the NNQE level (dotted line) does not depend on $\alpha$. 
}
\label{fig:msd4}
\end{figure}

In Fig.~\ref{fig:msd4},  we show the behavior of the MSD vs. time, for different values of $\alpha$, 
and  for $\xi=$ 0.08 (a) and 0.05 (b).
Besides the outcomes from the exact numerical integration (solid lines), we plot the approximate analytical expression valid for intermediate-long-times (dashed) and the analytical prediction of the NNQE (dotted), which will be presented below. 
The very good agreement between our analytical expressions and exact numerical results validates the formulas that we will derive in the next section using the approximate analytical expressions.

By means of the approximate solutions derived in 
Sec.~\ref{sec:sol-alfa}, 
valid for intermediate-long time scales (see derivation  in Appendix~\ref{app:theo}), 
we have
\begin{eqnarray} \nonumber
\langle x^2  \rangle_\alpha (t)
& \simeq &  \langle x^2 \rangle^{\textbf{C}} 
+ \langle x^2 \rangle_\alpha^{\textbf{T}}(t) \\
\label{eq:MSD}
&\simeq&  \langle x^2 \rangle^{\textbf{C}} + \frac{ 4t^{\frac{3\alpha}{2}}}{\mathcal{Z}\, \Gamma(\frac{3\alpha}{2}+1) }  
\,,
\end{eqnarray}
where, within the considered timescales, the second term is negligible compared to the 
first, which is time-independent.
Moreover,  $\langle x^2 \rangle^{\textbf{C}}$  is the same for any $\alpha$, hence equal to the expression already obtained for $\alpha=1$~\cite{Defaveri2020}, 
namely,  for any $\mu>1$,

\begin{equation} \label{eq:xc}
 \langle x^2 \rangle^{\textbf{C}}\simeq
 \frac{\int_0^\infty x^2(e^{-v(x)/\xi}-h(x))  dx}{\int_{0}^\infty (e^{-v(x)/\xi}-1) dx}\,,
\end{equation}
where $h(x)=\sum_{k=0}^{\lfloor 3/\mu\rfloor} (-1)^k [v(x)/\xi]^k/k!$~\cite{Anteneodo2021}. That is, $h(x)$ has only one term for $\mu=4$ and two terms for $\mu=2$.  This auxiliary function is responsible for regularizing the integrand, ensuring that it will reach a finite value.

Time-independence occurs, according to Eq.~(\ref{eq:MSD}), for timescales such that
\begin{eqnarray} \label{eq:t**} 
 t \ll  t^{**} \equiv \left\{ 
 \langle x^2 \rangle^{\textbf{C}}\,\frac{\mathcal{Z}}{4}\,
 \Gamma\Bigl(\frac{3\alpha}{2}+1 \Bigr) \right\}^{\frac{2}{3\alpha}}  
 \sim e^{ \frac{2}{3\alpha\xi}}, 
\end{eqnarray}
recalling that 
 both $\mathcal{Z}$ and $\langle x^2\rangle^{\bf C} $ 
depend on $\xi$ and $\mu$, but not on $\alpha$. 
This  inequality constrains the condition found in Eq.~(\ref{eq:t*}), as far as  $t^{**}<t^{*}$. Then, the lifetime of the quasi stationary MSD increases with $1/\xi$ as can be seen by comparing both panels of Fig.~\ref{fig:msd4}. 
It also increases with  $1/\alpha$, as  seen by comparing the different curves in each panel.  
In the limit $\alpha \to 0$, $t^{**}\sim e^{2/(3\xi\alpha)} \to \infty$, meaning that the stagnation tends to last forever. However, if $\alpha=0$, the NNQE is trivial,  as it simply means that the initial condition remains frozen, similarly to the limit of zero temperature $\xi$.

\begin{figure}[t!]  
\centering
\includegraphics[width=0.5\textwidth]{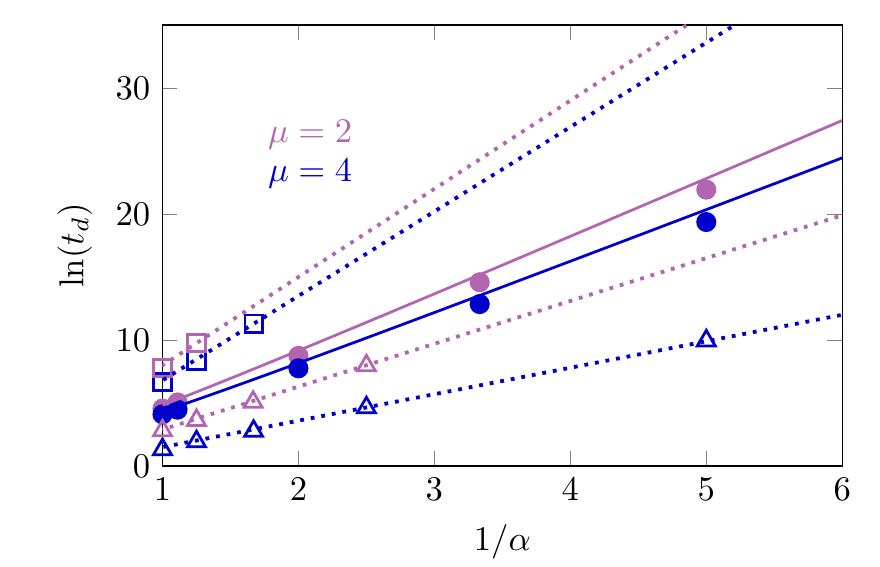}
\caption{ 
Duration $t_d$ of NNQE for the MSD (filled symbols) 
and for the energy (open  symbols).
The duration was obtained from the respective timeseries, by measuring the time at which the averaged observable departs from the NNQE value  by $\delta =0.1\%$. 
The full lines correspond to the estimates given by Eq.~(\ref{eq:td})  for the MSD and dotted lines to 
exponential fits for the energy data.  
The values of the relative temperature are
$\xi=0.05$ (circles),  $\xi=0.08$ (squares) and $\xi=0.1$ (triangles). 
Potentials with $\mu=2$ (light purple) and $\mu=4$ (dark blue) 
were considered.
 }
\label{fig:life}
\end{figure}

In Fig.~\ref{fig:life},  we display how the duration time $t_d$, during which  the MSD remains in NNQE, depends on $\alpha$. This duration  was estimated as the instant  at which the MSD exceeds the NNQE value by $\delta \%$ (symbols), where $\delta=0.1$. This estimate can be obtained by   identifying the last term in Eq.~(\ref{eq:MSD}) with the excess  $(\delta/100)\langle x^2\rangle^{\bf C} $,  which provides  the prediction

\begin{equation} \label{eq:td}
t_d
=t^{**} \left(
\delta/100\right)^\frac{2}{3\alpha} \sim (e^\frac{1}{\xi} \delta/100)^\frac{2}{3\alpha}.
\end{equation}
The effect of the shape of the potential is embodied in $t^{**}$ through $\mathcal{Z}$ and $\langle x^2\rangle^{\bf C}$. 
Notice in Fig.~\ref{fig:life} that $t_d$ increases exponentially with $1/\alpha$ following the predicted law.

It is important to remark that we are considering  the nondimensionalized variables, while in the 
transformation
$t/t_0 \to t$, the scaling factor $t_0= (x_0^2/\mathcal{K}_{\alpha})^{1/\alpha}$, 
in Eq.~(\ref{eq:scaling}), depends on $\alpha$. Hence, the observation that decreasing $\alpha$ prolongs the duration of NNQE, as observed for the scaled time (equivalent to setting $t_0=1$), is still true for the real time if 
$x_0^2/\mathcal{K}_{\alpha}\gtrsim 1$, but it can be inverted otherwise.  
In fact, to recover the original (nonscaled) variables, 
the time in Eqs.~(\ref{eq:t*}) and 
~(\ref{eq:t**}) must be multiplied by 
 $t_0=(x_0^2/{\cal K}_\alpha)^{1/\alpha}$.  
Therefore, if $x_0^2/{\cal K}_\alpha \lesssim \exp[-2/(3\xi)]$, 
 which can occur for relatively narrow potential well or large diffusivity coefficient, the lifetime of the MSD increases with $\alpha$.

\subsection{Energy}
\label{sec:energy}

\begin{figure}[t!]  
\centering
\includegraphics[width=0.5\textwidth]{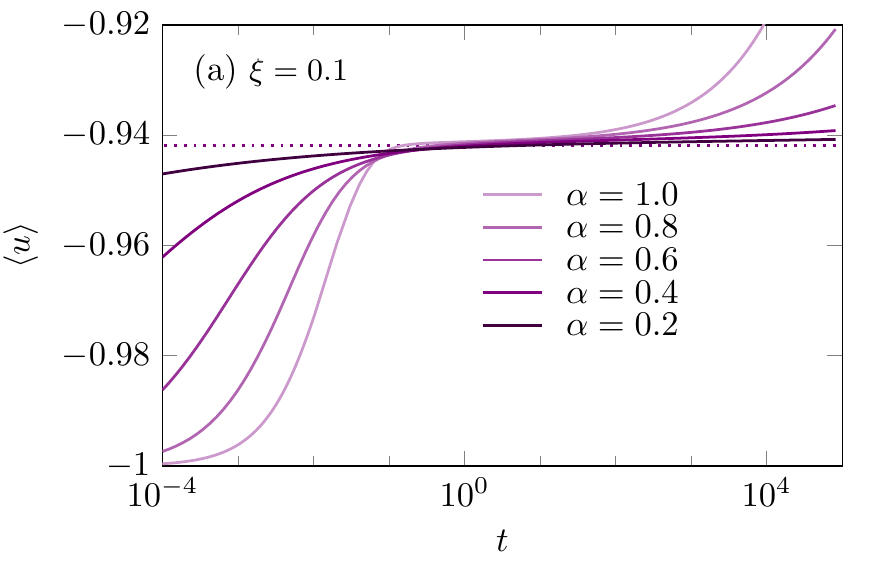} 
\includegraphics[width=0.5\textwidth]{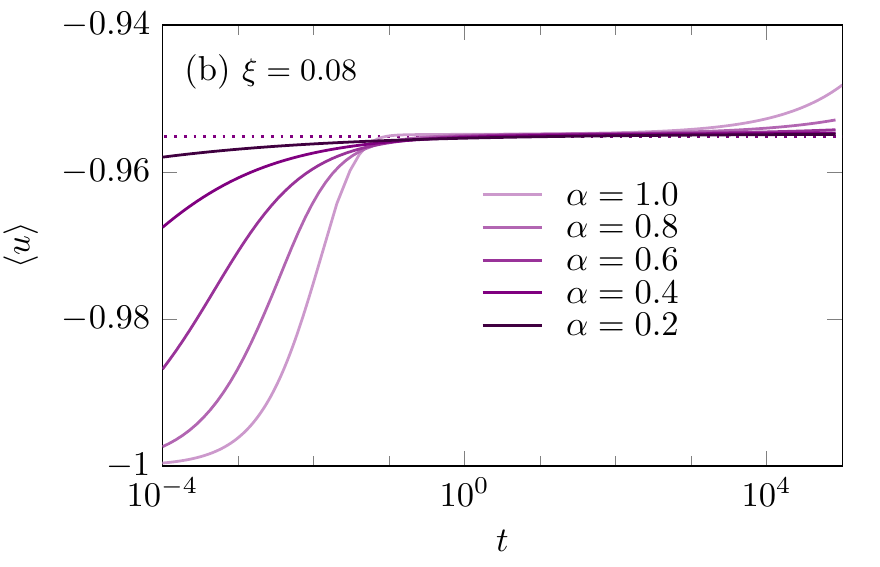} 
\caption{
Average energy vs. time,  for $\mu=4$,  
and different values of $\alpha$,  when $\xi=0.1$ (a) and $\xi=0.08$ (b). 
The  solid lines correspond to the exact numerical solution. 
The NNQE level (dotted line),  given by Eq.~(\ref{eq:NQE-E}), does not depend on $\alpha$. 
}
\label{fig:energia}
\end{figure}

In this section we discuss the time behavior and NNQE duration for a thermodynamical observable, the energy. 
The time evolution of the average energy is shown in Fig.~\ref{fig:energia}, for $\xi=0.1$ (a) and 0.08 (b),  with different values of $\alpha$. 
Also in this case, the lifespan increases, diverging exponentially with $1/\alpha$, as shown in Fig.~\ref{fig:life}.

Since the energy vanishes  asymptotically, the tails have a smaller effect on the average energy than they do for the moments of the PDF. 
As a first consequence, the plateau of the energy can emerge for higher relative temperature $\xi$ than for the MSD. Note in Fig.~\ref{fig:energia}, where we plot the average energy versus time, that a plateau appears for $\xi=0.1$, 
while  not  for the MSD (see Fig.~\ref{fig:msd4}).
Moreover, the lifetime for the average energy is much larger than for the MSD at the same relative temperature.
This is because the MSD is more sensitive to the large $x$ behavior of the density than the energy observable.
The NNQE value of the energy can be predicted by averaging the potential energy with the regularized procedure, yielding for $\mu>1$~\cite{Defaveri2020}
\begin{equation} \label{eq:NQE-E}
 \langle u \rangle^{\textbf{C}}\simeq
 \frac{\int_0^\infty v(x) e^{-v(x)/\xi}   dx}{\int_{0}^\infty (e^{-v(x)/\xi}-1) dx}\,,
\end{equation} 
represented by dotted lines in Fig.~\ref{fig:energia}, in good agreement with the corresponding numerical solutions.  
As another consequence of the asymptotic behavior of the observable,   only  the denominator needs to be regularized.

Through an heuristic reasoning, we can estimate that the correction to the stagnation value, 
$\langle u \rangle^{\textbf{T}} (t)$, which  describes the departure from the plateau,  
is 
${\cal O}(t^{\alpha/2})$, the same law as in Eq.~(\ref{eq:tail-correction}), which will lead to a departure time following a law similar to Eq.~(\ref{eq:t**}). In the central region, the PDF is proportional to the BG factor $e^{-v_\mu(x)/\xi}$, and since the energy only shows non-negligible values in this central region, the approximation is expected to hold as long as our approximation for the PDF holds. The departure times $t_d$, observed numerically, for the energy are plotted in Fig.~\ref{fig:life} (open symbols), showing exponential dependence with $1/\alpha$, as for the MSD.

\section{Discussion and final remarks}
\label{sec:final}

We have introduced a fractional time derivative of Caputo type in the FPE  (actually, a Riemann Liouville integral equation)  for an asymptotically flat potential with a well at the origin. 
We have shown that this potential leads to NNQE states, 
in which the averaged  dynamical and thermodynamical  observables can be obtained by a regularized procedure of BG statistics, valid for the whole range of anomalous processes, $0<\alpha\le 1$, since this index acts only on the time domain and not in the spatial one.
The reason for this universal behavior is that the quasi-equilibrium is controlled by the BG factor, and not by $\alpha$, as the latter is responsible for dynamical effects only. 
Hence, in that sense, the regularized approach obtained for the integer order case~\cite{Defaveri2020,Anteneodo2021} is general.

With regard to the lifetime of  the NNQE states for the MSD, 
for fixed $\alpha$, it increases exponentially with  the  ratio  between the potential depth at the origin and the temperature,  $1/\xi$,  as $\exp[2/(3\alpha\xi)]$, as given by Eq.\,(\ref{eq:t**}). 
The fractional exponent $\alpha$ contributes to this exponential similarly to  $\xi$. The lifetime is typically longer the more subdiffusive the dynamics (the smaller $\alpha$). However, the opposite may happen, in the original (non-scaled) variable,  when the width of the potential $x_0$  is sufficiently small or  the diffusivity constant ${\cal K}_\alpha$ sufficiently large, as discussed in Sec.~\ref{sec:times}.

Looking forward, it would be interesting to explore other generalization of the FPE, with both time and space fractional derivatives. It may be also worth considering systems driven by non-Gaussian and colored heat baths, as well as the behavior at very long times, after the NNQE regime, once enough particles have escaped the well.
 In the limit of large $\xi$, that is, $V_0 \gtrsim K_B T$, the particles will escape and return to the origin many times. This represents another direction of research, where we expect the statistics to be also related to the Boltzmann-Gibbs distribution, and 
infinite ergodic theory to hold~\cite{Aghion2020}.  Finally, ergodic properties of time averages and their relation to the Boltzmann-Gibbs measure could be studied as well.\\[1cm]

{\bf ACKNOWLEDGMENTS:}
We acknowledge partial financial support from CAPES (code 001), CNPq and FAPERJ, in Brazil. The support of Israel Science Foundation's grant 1614/21 is acknowledged.


\setcounter{figure}{0}
\renewcommand{\thefigure}{A\arabic{figure}}

\appendix 

\section{Theoretical averages}
\label{app:theo}

  For sufficiently large $t$, the part of the solution $ P^{\text{{\bf C}}}(x,t)$ becomes time-independent, hence, Eq.~(\ref{eq:P1alfa}) captures the behavior of the majority of the particles. Then, when the system is in the intermediate long timescale, we can determine the MSD by the procedure that will be detailed below. We will omit the subindex $\alpha$, to simplify the notation.

To calculate the average of an observable, we split the integration as follows 
\begin{eqnarray} \nonumber
 \langle \mathcal{O} \rangle \simeq  2\int_{0}^{\ell} \mathcal{O}(x) P^{\text{{\bf C}}}(x,t) dx + 2\int_{\ell}^{\infty} \mathcal{O}(x) P^{\text{{\bf T}}}(x,t) dx. 
\end{eqnarray}

First note that the first integral is nearly constant in time. 
The integral corresponding to the region of the tails, $\ell \leq x $, can be calculated via the Laplace transform in the time variable.

For the MSD, we obtain 
\begin{eqnarray}
\widetilde{\langle x^2 \rangle}^{\textbf{T}} (s) & \simeq&  2 \int_{\ell}^{\infty} x^2 \widetilde{P}^{\textbf{T}} (x,s) dx  \\
& \simeq &   \frac{2}{  \mathcal{Z} s }\int_{\ell}^{\infty}  x^2   e^{-s^{\frac{\alpha}{2}}|x|}  dx \nonumber \\
&= & \left(  \frac{\ell^2}{2 s^{\frac{\alpha}{2}+1}} + \frac{\ell}{ s^{\alpha+1}}  + \frac{1}{  s^{\frac{3\alpha}{2}+1}}  \right) \frac{ 4 }{ \mathcal{Z}} e^{-s^{\frac{\alpha}{2}}\ell}  \,, \nonumber
\end{eqnarray}
which, for long time, i.e., $s\to 0$, becomes
\begin{eqnarray}
\widetilde{\langle x^2 \rangle}^{\textbf{T}} (s)  
& \sim &   
  \frac{4}{\mathcal{Z}\,s^{\frac{3\alpha}{2}+1} } \,,   
\end{eqnarray}
and after applying the inverse Laplace transform, we arrive at
\begin{eqnarray}
\langle x^2 \rangle^{\textbf{T}} (t) &\simeq & 
\frac{4}{\mathcal{Z}\,\Gamma(\frac{3\alpha}{2}+1) }  t^{\frac{3\alpha}{2}}  +{\cal O}(t^\frac{\alpha}{2})\,.  
\end{eqnarray}
That is, in the  large-$t$ limit, we obtain
\begin{eqnarray} \label{eq:scalingMSD}
\langle x^2  \rangle(t) & \simeq & \langle x^2 \rangle^{\textbf{C}} + \langle x^2 \rangle^{\textbf{T}}(t) \nonumber  \\ & \simeq & \langle x^2 \rangle^{\textbf{C}} +  \frac{4}{\mathcal{Z}\, \Gamma(\frac{3\alpha}{2}+1) }  t^{\frac{3\alpha}{2}}
\,.
\end{eqnarray}



\begin{thebibliography}{4}

\bibitem{risken}
H. Risken, 
{\it The Fokker-Planck Equation}, 
\href{https://doi.org/10.1007/978-3-642-61544-3}{Springer, Berlin}, (1996).


\bibitem{fermi} 
 E. Fermi, 
 {\it On the probability of the quantum states}, 
 \href{https://doi.org/10.1007/BF01327311}{Z. Phys. 26, 54-56} (1924).

\bibitem{plastino}
 A. Plastino, M. C. Rocca, and G. L. Ferri, 
 {\it Resolving the partition function’s paradox of the hydrogen atom}, 
 \href{https://doi.org/10.1016/j.physa.2019.122169}{Physica A 534, 122169} (2019).

\bibitem{Defaveri2020}
L. Defaveri, C. Anteneodo, D. A. Kessler, E. Barkai, 
{\it Regularized Boltzmann-Gibbs statistics for a Brownian particle in a nonconfining field}, 
\href{https://doi.org/10.1103/PhysRevResearch.2.043088}{Phys. Rev. Research 2 043088} (2020).

\bibitem{Anteneodo2021}
C. Anteneodo, L. Defaveri, E. Barkai, D. A. Kessler,
{\it Non-Normalizable Quasi-Equilibrium Solution of the Fokker–Planck Equation for Nonconfining Fields}, 
\href{https://doi.org/10.3390/e23020131}{Entropy 23(2): 131} (2021).


\bibitem{bertin2003}
E. M. Bertin and J.-P. Bouchaud, 
{\it Subdiffusion and localization in the one-dimensional trap model}, 
\href{https://doi.org/10.1103/PhysRevE.67.026128}{Phys. Rev. E 67, 026128} (2003).

\bibitem{sollich2005}
P. Mayer and P. Sollich, 
{\it Observable dependent quasiequilibrium in slow dynamics},
\href{https://doi.org/10.1103/PhysRevE.71.046113}{Phys. Rev. E 71, 046113} (2005). 


\bibitem{Oliveira2019}
F. A. Oliveira, R. M. Ferreira, L. C. Lapas, M. H. Vainstein, 
{\it Anomalous diffusion: A basic mechanism for the evolution of inhomogeneous systems}, 
\href{https://doi.org/10.3389/fphy.2019.00018}{Frontiers in Physics, 7, 18} (2019).

\bibitem{Bouchaud1990}
J. P. Bouchaud, A. Georges, 
{\it Anomalous diffusion in disordered media: statistical mechanisms, models and physical applications},  
\href{https://doi.org/10.1016/0370-1573(90)90099-N}{Phys.  Rep. 195(4-5), 127-293} (1990).


\bibitem{bellour2002}
M. Bellour, M. Skouri, J.-P. Munch,  P. H\'ebraud, 
{\it Brownian motion of particles embedded in a solution of giant micelles},
\href{https://doi.org/10.1140/epje/i2002-10026-0}{Eur. Phys. J. E 8, 431–436} (2002).

\bibitem{castillo2008}
J. Galvan-Miyoshi, J.  Delgado, R. Castillo,
{\it Diffusing wave spectroscopy in Maxwellian fluids}, 
\href{https://doi.org/10.1140/epje/i2007-10335-8}{Eur. Phys. J. E 26, 369–377} (2008). 

\bibitem{jeon2013}
J.-H. Jeon, N. Leijnse, L.B. Oddershede, R. Metzler, 
{\it Anomalous diffusion and power-law relaxation of the time averaged mean squared displacement in worm-like micellar solutions}, 
\href{http://doi.org/10.1088/1367-2630/15/4/045011}{New J. Phys. 15 045011} (2013).

\bibitem{Corci2021}
C. {\AA}berg,  B. Poolman,
{\it Glass-like characteristics of intracellular motion in
human cells}
\href{https://doi.org/10.1016/j.bpj.2021.04.011}{Biophysical Journal 120, 2355-2366} (2021).

\bibitem{Corci2023}
B. Corci,   O. Hooiveld,   A. Dolga  and  C. {\AA}berg  ,
{\it Extending the analogy between intracellular motion in mammalian cells and glassy dynamics}
\href{https://doi.org/10.1039/D2SM01672A}{Soft Matter, Accepted Manuscript,} (2023).

\bibitem{Kurilovich2020}
A. A. Kurilovich, V. V. Palyulin, V.N. Mantsevich, K. J. Stevenson, A. V. Chechkincd, V. V. Palyulin,
{\it Complex diffusion-based kinetics of photoluminescence in semiconductor nanoplatelets}
\href{https://doi.org/10.1039/D0CP03744C}{Phys. Chem. Chem. Phys. 22, 24686} (2020).


\bibitem{Kurilovich2022}
A. A. Kurilovich, K. J. Stevenson, V. N. Mantsevich, A. V. Chechkin, Y. Mardoukhi, V. V. Palyulin,
{\it Non-Markovian diffusion of excitons in layered perovskites and transition metal dichalcogenides}
\href{https://doi.org/10.1039/d2cp00557c}{Phys. Chem. Chem. Phys. 24, 13941} (2022).

\bibitem{Vilk2022}
O. Vilk, E. Aghion, T. Avgar, C. Beta, O. Nagel, A. Sabri,  R. Sarfati, D.K. Schwartz, M. Weiss, D. Krapf, R. Nathan,
{\it Unravelling the origins of anomalous diffusion: from molecules to migrating storks}, 
\href{https://doi.org/10.1103/PhysRevResearch.4.033055}{Phys. Rev. Research 4, 033055} (2022).

\bibitem{Munoz2021}
G. Mu\~noz-Gil, G. Volpe, M.A. Garcia-March,  et al.,  
{\it Objective comparison of methods to decode anomalous diffusion},  
\href{https://doi.org/10.1038/s41467-021-26320-w}{Nat. Commun. 12 6253} (2021).

\bibitem{Tsallis1996}
C. Tsallis,  D. J. Bukman,
{\it Anomalous diffusion in the presence of external forces: Exact time-dependent solutions and their thermostatistical basis}, 
\href{https://doi.org/10.1103/PhysRevE.54.R2197}{Phys. Rev. E, 54, R2197(R)} (1996).

\bibitem{escape2001}
E.K. Lenzi, C. Anteneodo, L. Borland
{\it Escape time in anomalous diffusive media}, 
\href{https://journals.aps.org/pre/abstract/10.1103/PhysRevE.63.051109}{Phys. Rev. E 63, 051109} (2001).


\bibitem{nonlinear-book}
T.D. Frank, 
{\it Nonlinear Fokker-Planck Equations: Fundamentals and Applications} (Springer-Verlag, Berlin, 2005).

\bibitem{MandelbrotvanNess}
B. B. Mandelbrot, J. W. van Ness, 
{\it  Fractional Brownian Motions, Fractional Noises and Applications},
\href{https://doi.org/10.1137/1010093}{SIAM Review, 10, 422-437} (1968).

\bibitem{Metzler1999}
R. Metzler, E. Barkai, J. Klafter, 
{\it Anomalous diffusion and relaxation close to thermal equilibrium: A fractional Fokker-Planck equation approach},
\href{https://doi.org/10.1103/PhysRevLett.82.3563}{Phys. Rev. Lett. 82, 3563} (1999).

\bibitem{Metzler2000}
R. Metzler, J.  Klafter, 
{\it The random walk's guide to anomalous diffusion: a fractional dynamics approach},
\href{https://doi.org/10.1016/S0370-1573(00)00070-3}{Phys. Rep. 339(1), 1-77} (2000).


\bibitem{Sibatov2013}
V. Uchaikin  and R. Sibatov,
{\it  Fractional Kinetics in Solids
Anomalous Charge Transport in Semiconductors, Dielectrics and Nanosystems}
 \href{https://doi.org/10.1142/8185 }{World Scientific} (2013).

\bibitem{Sibatov2020}
R.T. Sibatov and H.Sun, 
{\it Dispersive Transport Described by the Generalized
Fick Law with Different Fractional Operators}
 \href{https://www.mdpi.com/2504-3110/4/3/42}{Fractal Fract. 4, 42} (2020).
 
\bibitem{Henry2010}
B. I. Henry, T. A. M. Langlands, and P. Straka,
{\it Fractional Fokker-Planck Equations for Subdiffusion with Space- and Time-Dependent Forces},
\href{https://doi.org/10.1103/PhysRevLett.105.170602}{Phys. Rev. Lett. 105, 170602} (2010).

\bibitem{Magdziarz2007}
M. Magdziarz, A. Weron, and K. Weron,
{\it Fractional Fokker-Planck dynamics: Stochastic representation and computer simulation},
\href{https://doi.org/10.1103/PhysRevE.75.016708}{Phys. Rev. E 75, 016708} (2007)

\bibitem{Metzler2022}
R. Metzler, A. Rajyaguru and B. Berkowitz,
{\it Modelling anomalous diffusion in semi-infinite disordered systems and porous media},
\href{htpps://doi.org/10.1088/1367-2630/aca70c}{New J. Phys. 24 123004} (2022)

\bibitem{David2022}
S. A. David, C. A. Valentim, and A. Debbouche,
{\it Fractional Modeling Applied to the Dynamics of the Action Potential in Cardiac Tissue},
\href{https://doi.org/10.3390/fractalfract6030149}{Fractal Fract.  6(3), 149} (2022)


\bibitem{Evangelista-book}
L. R. Evangelista, E. K. Lenzi,
{\it An Introduction to Anomalous Diffusion and Relaxation},
\href{https://doi.org/10.1007/978-3-031-18150-4}{Springer Cham, Torino} (2023).

\bibitem{Sokolov2001}
I. M. Sokolov, 
{\it Thermodynamics and fractional Fokker-Planck equations}, 
\href{https://doi.org/10.1103/PhysRevE.63.056111}{Phys. Rev. E, 63, 056111} (2001).

\bibitem{Podlubny1999}
I. Podlubny, 
{\it Fractional differential equations},  Mathematics in science and engineering, 198, 41-119 (1999).

\bibitem{Barkai2000}
E. Barkai,   R. Metzler,  J. Klafter, 
{\it From continuous time random walks to the fractional Fokker-Planck equation}, 
\href{https://doi.org/10.1103/PhysRevE.61.132}{Phys. Rev. E, 61, 132} (2000).



\bibitem{Dybiec2015}
B. Dybiec, I.M. Sokolov, 
{\it Estimation of the smallest eigenvalue in fractional escape problems: Semi-analytics and fits}, 
\href{https://doi.org/10.1016/j.cpc.2014.10.007}{Comput. Phys. Commun., 187, 29-37} (2015).

\bibitem{Sliusarenko2010}
O. Y. Sliusarenko,  V. Y. Gonchar, A. V. Chechkin, I. M. Sokolov, R. Metzler,  
{\it Kramers-like escape driven by fractional Gaussian noise}, 
\href{https://doi.org/10.1103/PhysRevE.81.041119}{Phys. Rev. E, 81, 041119}  (2010).


\bibitem{MetzlerKlafter2000}
R. Metzler, J. Klafter, 
{\it Boundary value problems for fractional diffusion equations}. 
\href{https://doi.org/10.1016/S0378-4371(99)00503-8}{Physica A: Statistical Mechanics and its Applications, 278(1-2), 107-125}  (2000).


 
\bibitem{Gorenflo2007}
R. Gorenflo, F. Mainardi, A. Vivoli, 
{\it Continuous-time random walk and parametric subordination in fractional diffusion},
\href{
https://doi.org/10.1016/j.chaos.2007.01.052}{
Chaos, Solitons and Fractals, 34,89-103} (2007).

\bibitem{Weron2010} 
K. Weron, A. Jurlewicz, M. Magdziarz, A. Weron, and J. Trzmiel,
{\it  Overshooting and undershooting subordination scenario for fractional two-power-law relaxation responses},
\href{https://doi.org/10.1103/PhysRevE.81.041123}{Phys. Rev. E 81, 041123} (2010)

\bibitem{Ewa2010}
B. Dybiec, E. Gudowska-Nowak, 
{\it Subordinated diffusion and CTRW asymptotics},
\href{https://doi.org/10.1063/1.3522761}{Chaos 20, 043129} (2010).

 \bibitem{Wang2020}
W. Wang, and E. Barkai,
{\it Fractional diffusion-advection-asymmetry  equation},
\href{https://doi.org/10.1103/PhysRevLett.125.240606}{Phys. Rev. Lett.  125, 240606}
{Editor's suggestion} (2020).

 \bibitem{Chechkin2021}
A. Chechkin, and I.M. Sokolov, 
{\it Relation between generalized diffusion equations and subordination schemes},
\href{}{Phys. Rev.E  103, 032133} (2021).

\bibitem{Zhou2022}
T. Zhou, P. Trajanovski, P. Xu, W. Deng, T. Sandev and L. Kocarev,
{\it Generalized diffusion and random search processes},
\href{https://doi.org/10.1088/1742-5468/ac841e}{J. Stat. Mech. 093201} (2022).

\bibitem{Wang2022}
X. Wang and Y. Chen
{\it Ergodic property of random diffusivity system with trapping events},
\href{https://doi.org/10.1103/PhysRevE.105.014106}{Phys. Rev. E 105, 014106} (2022).

 
 \bibitem{Barkai2001}
E. Barkai, 
{\it Fractional Fokker-Planck equation, solution, and applications}, 
\href{https://doi.org/10.1103/PhysRevE.63.046118}{Phys. Rev. E 63, 046118} (2001).

\bibitem{penson}
K. A. Penson and K. G\'orska, 
{\it Exact and Explicit Probability Densities for One-Sided L\'evy Stable Distributions}, 
\href{https://doi.org/10.1103/PhysRevLett.105.210604}{Phys. Rev. Lett. 105, 210604}  (2010).



\bibitem{Deng2007}
W. Deng,
{\it Numerical algorithm for the time fractional Fokker–Planck equation},
\href{https://doi.org/10.1016/j.jcp.2007.09.015}{Journal of computational physics 2, 1510-1522} (2007).

\bibitem{Deng2009}
W. Deng,
{\it Finite Element Method for the Space and Time Fractional Fokker–Planck Equation},
\href{https://doi.org/10.1137/080714130}{{\it SIAM} J. Math. Anal., 47 (1986), pp. 204-226} (2009)

 
\bibitem{Bel2006}
G. Bel, E. Barkai,
{\it Random Walk to a Nonergodic Equilibrium Concept}
 \href{https://doi.org/10.1103/PhysRevE.73.016125}{Phys. Rev. E  73,  016125} (2006).
 


\bibitem{Aghion2020}
E. Aghion, D. A. Kessler, E. Barkai,
{\it Infinite ergodic theory meets Boltzmann statistics}, 
\href{https://doi.org/10.1016/j.chaos.2020.109890}{Chaos, Solitons \& Fractals, 138, 109890} (2020).



\end{thebibliography}
\end{document}